\documentclass[aps,prl,twocolumn,groupedaddress,draft,showpacs,intlimits,amsmath,amssymb,floats, superscriptaddress]{revtex4}
\usepackage{bm,times}
\usepackage[final]{graphicx}
\usepackage{epsfig}
\usepackage{color}
\definecolor{DarkBlue}{rgb}{0.1,0.1,0.5}
\definecolor{Red}{rgb}{0.9,0.0,0.1}
\definecolor{Green}{rgb}{0.0,0.99,0.0}

\begin{document}

\title{Finite temperature spin-dynamics and phase transitions in spin-orbital models}

\author{C.-C. Chen}
\affiliation{Department of Physics and Geballe Laboratory for Advanced Materials, Stanford University, CA 94305, USA}
\affiliation{Stanford Institute for Materials and Energy Science, SLAC National Accelerator Laboratory, 2575 Sand Hill Road, Menlo Park, CA 94025, USA}
\author{B. Moritz}
\affiliation{Stanford Institute for Materials and Energy Science, SLAC National Accelerator Laboratory, 2575 Sand Hill Road, Menlo Park, CA 94025, USA}
\author{J. van den Brink}
\affiliation{Stanford Institute for Materials and Energy Science, SLAC National Accelerator Laboratory, 2575 Sand Hill Road, Menlo Park, CA 94025, USA}
\affiliation{Institute Lorentz for Theoretical Physics, Leiden University, P.O. Box 9506, 2300 RA Leiden, The Netherlands}
\author{T. P. Devereaux}
\affiliation{Department of Physics and Geballe Laboratory for Advanced Materials, Stanford University, CA 94305, USA}
\affiliation{Stanford Institute for Materials and Energy Science, SLAC National Accelerator Laboratory, 2575 Sand Hill Road, Menlo Park, CA 94025, USA}
\author{R. R. P. Singh}
\email[Corresponding author:\,]{singh@physics.ucdavis.edu}
\affiliation{Department of Physics, University of California, Davis, CA 95616, USA}

\date{\today}

\begin{abstract}
We study finite temperature properties of a generic spin-orbital model relevant to transition metal compounds, having coupled quantum Heisenberg-spin and Ising-orbital degrees of freedom. The model system undergoes a phase transition, consistent with that of a 2D Ising model, to an orbitally ordered state at a temperature set by short-range magnetic order. At low temperatures the orbital degrees of freedom freeze-out and the model maps onto a quantum Heisenberg model. The onset of orbital excitations causes a rapid scrambling of the spin spectral weight away from coherent spin-waves, which leads to a sharp increase in uniform magnetic susceptibility just below the phase transition, reminiscent of the observed behavior in the Fe-pnictide materials.
\end{abstract}

\maketitle

Correlated materials exhibit intriguing phenomena arising from the interplay between spin, charge, lattice, and orbital degrees of freedom. Orbital degrees of freedom can emerge in multi-band systems such as $3d$ transition metal compounds. In these systems, spins and orbitals are strongly coupled as spin exchange is the dominant interaction between different orbital occupations, which in turn support different spin order. This correlation can lead to a phase transition in one or both variables, the collective effects of which can be antecedent or subsequent to a lattice structural transition \cite{Kugel}. A paradigmatic example is manganites where orbital ordering is essential in explaining the magnetic properties and phase transitions \cite{Tokura}. 

\begin{figure}[t!]
\includegraphics[width=\columnwidth]{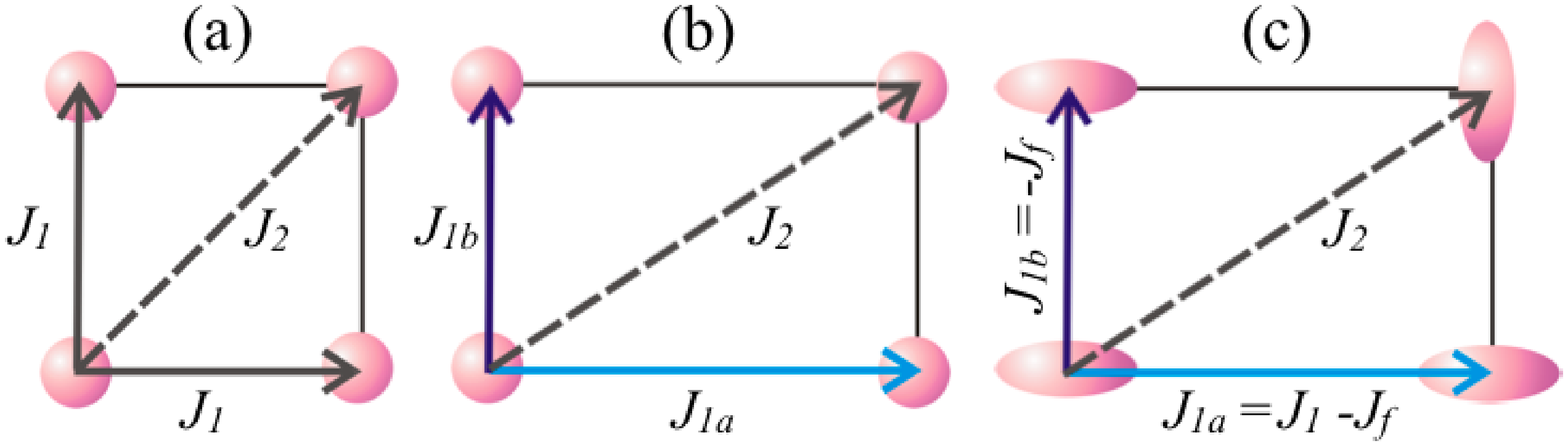}
\includegraphics[width=\columnwidth]{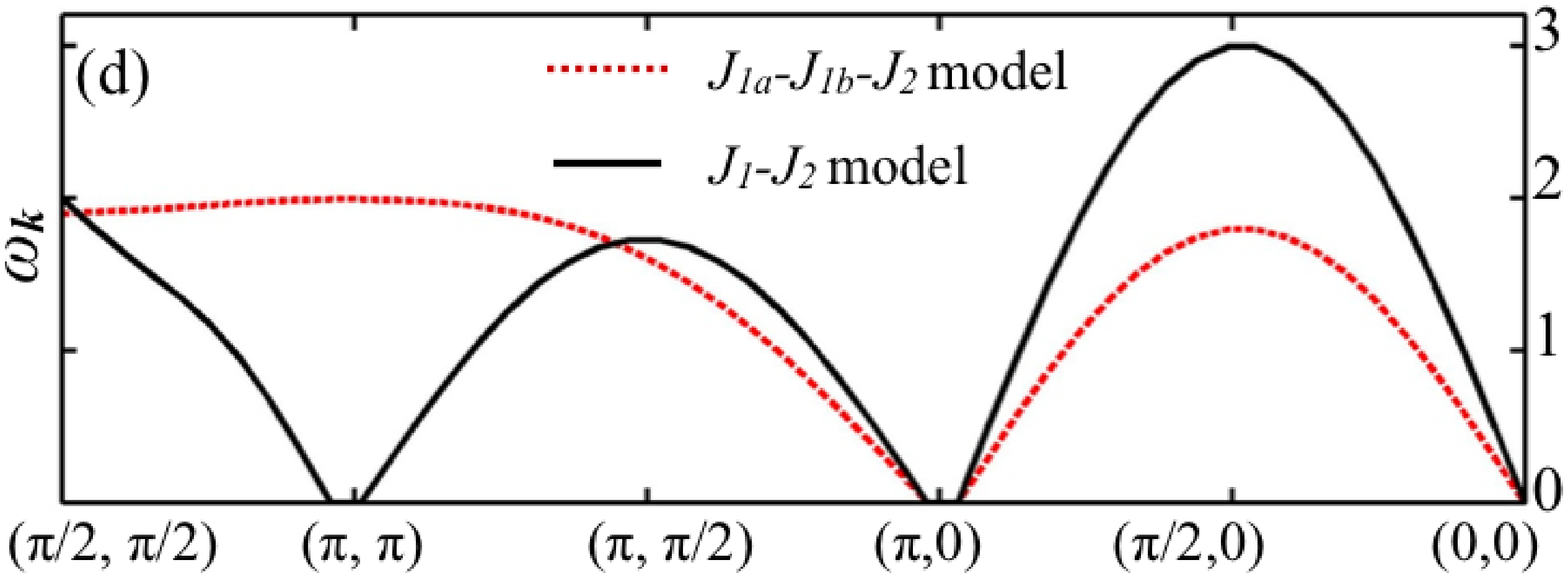}
\caption{
(Color online) Schematic representation of various Hamiltonians: (a) The $J_1$-$J_2$, (b) the $J_{1a}$-$J_{1b}$-$J_2$, and (c) the spin-orbital models. At zero temperature the additional orbital degrees of freedom freeze-out, and the spin-orbital model reduces to the $J_{1a}$-$J_{1b}$-$J_2$ model. (d) The magnon dispersion $\omega_{\textbf{k}}$ (in units of $J_1/J_{1a}$) calculated from linear spin wave theory. Here $J_{1b}=-0.1J_{1a}$, $J_2=0.4J_{1a}$ in the $J_{1a}$-$J_{1b}$-$J_2$ model, and $J_2=J_1$ in the $J_1$-$J_2$ model. The spin wave energy forms a maximum at $(\pi,\pi)$ in the $J_{1a}$-$J_{1b}$-$J_2$ model, which is a minimum in the $J_1$-$J_2$ model.
}\label{fig:LSWT}
\end{figure}

The newly discovered Fe-pnictide superconductors \cite{pnictide_family} display superconductivity in close proximity to magnetic order. The observed collinear $(\pi,0)$ magnetic order \cite{Dai} has been studied theoretically from both weak- \cite{weak} and strong-coupling points of view \cite{strong}. In particular, an anti-ferromagnetic (AF) coupled $J_1$-$J_2$ Heisenberg model on a 2D square lattice (depicted in Fig. \ref{fig:LSWT} (a)) can give rise to an AF $(\pi,0)$ order when $J_2\geq J_1/2$ \cite{J1_J2}. Alternatively, this $(\pi,0)$ order may be obtained through an anisotropic $J_{1a}$-$J_{1b}$-$J_2$ model \cite{savrasov}, where one has strong AF coupling in the $x$-direction, and ferromagnetic coupling along the $y$-direction, as shown Fig. \ref{fig:LSWT} (b). Interestingly, recent neutron scattering data \cite{Neutron_SW} indicate that the magnon energy is a maximum at momentum transfer $(\pi,\pi)$. This strongly favors the $J_{1a}$-$J_{1b}$-$J_2$ scenario which reproduces the observed spin wave dispersion, see Fig. \ref{fig:LSWT} (d).

A possible microscopic origin for the anisotropy in the $J_{1a}$-$J_{1b}$-$J_2$ model is orbital ordering \cite{Jeroen_OO, Rajiv_OO}. When the orbitals are ordered, the lattice distorts and the orbital lobe orientations can cause a vanishing effective hopping in certain directions, as in 1D edge-sharing copper oxides \cite{edge-sharing}. In conjunction with double exchange \cite{Double_Exchange}, orbital ordering, affecting the direct and super-exchange processes, can lead to even sign-changing anisotropic exchange interactions. Proposals have been put forth that consider ordering between the Fe 3$d_{xz}$ and 3$d_{yz}$ orbitals as a possible mechanism for the observed magnetism of the Fe pnictides \cite{Jeroen_OO, Rajiv_OO, other_OO}. These proposals remain controversial in part because early band structure calculations \cite{savrasov}, which agree well with a variety of experiments, show a very small difference in the occupation of $d_{xz}$ and $d_{yz}$ orbitals in the magnetically ordered tetragonal calculation \cite{savrasov-privatecomm}. On the other hand, recent \emph{ab initio} calculations suggest robust orbital order using Wannier orbitals \cite{WeiKu_OO}. Indeed, if the magnon energy is a maximum at ($\pi,\pi$) as reported in Ref. \cite{Neutron_SW}, this implies not just a small anisotropy due to for example structural considerations, but an extreme sign-changing one associated with additional broken symmetry \cite{Rajiv_OO}.

In this paper, we address the question: If the anisotropy in exchange constants observed in neutron scattering is related to orbital order, what other consequences follow. To answer this, we consider the following spin-orbital Hamiltonian relevant to the schematic  in Fig. \ref{fig:LSWT} (c):
\begin{eqnarray}
H&=& J_1\sum_i [\mathbf{S}_i\cdot\mathbf{S}_{i+\hat x} ~ n_i n_{i+\hat x} +\mathbf{S}_i\cdot \mathbf{S}_{i+\hat y} ~ (1-n_i)(1-n_{i+\hat y})]
\nonumber
\\
&-& \frac{J_f}{2}\sum_{<ij>} \mathbf{S}_i\cdot\mathbf{S}_j+\frac{J_2}{2}\sum_{\ll ij\gg} \mathbf{S}_i\cdot\mathbf{S}_j,
\end{eqnarray}
where $n_i$ is an Ising variable taking values 0 or 1, and $\mathbf{S}_i$ is a spin-$\frac{1}{2}$ operator: $\mathbf{S}_i\cdot\mathbf{S}_i=S(S+1)$, and $S=\frac{1}{2}$. The sums $<>$ and $\ll\gg$ run over nearest- and second nearest-neighbors, respectively. This model describes a system consisting of two orbitals per site, with the occupation controlled by the Ising variables: $n_i=0$ represents orbital 1 occupied; $n_i=1$ represents orbital 2 occupied.

When the interactions are dominated by an AF coupled $J_1$, the above model finds its lowest energy configuration in a perfect ferro-orbitally ordered state corresponding to all $n_i=0$ or 1. Therefore, at zero temperature $T=0$ this Hamiltonian reduces to the $J_{1a}$-$J_{1b}$-$J_2$ model with $J_{1a}=J_1-J_f$ and $J_{1b}=-J_f$. On the other hand, the finite temperature properties would be quite different due to orbital fluctuations and excitations. In the following calculations we take the parameters from neutron scattering data on CaFe$_2$As$_2$ \cite{Neutron_SW}: $SJ_{1a}= 50$ meV, $SJ_f=6$ meV, and $SJ_2=20$ meV.

We are interested in the finite temperature spin dynamics of these systems; however, there are few numerical methods capable of accomplishing this in a controlled manner. We use the exact diagonalization (ED) technique, which has been utilized extensively to investigate both zero and finite temperature properties for various quantum lattice models \cite{Dagotto}. We use $N$=16 site square plaquettes, already requiring a large computational effort due to the additional orbital degrees of freedom. Lattice translation, rotation, reflection and Ising-orbital inversion symmetries reduce the $2^{16}$ Ising configurations to 733 distinct ones. We \emph{fully} diagonalize the Hamiltonian in these Ising sectors and calculate dynamical quantities.

Our main results are as follows: (i) In a purely 2D system, where in accord with the Mermin-Wagner theorem the spin-rotational symmetry can not be spontaneously broken except at $T=0$, the orbital degrees of freedom undergo a phase transition at a temperature scale $\sim0.2 J_1$ set by short range magnetic order. (ii) At temperatures below $0.1 J_1$, the Ising variables are completely frozen and the model maps onto the $J_{1a}$-$J_{1b}$-$J_2$ model. Above $T=0.1 J_1$, the onset of orbital excitations causes a scrambling of the spin spectral weight, leading to sharply diminished spin-wave peaks. (iii) There is a sudden increase in the uniform magnetic susceptibility just below the phase transition. Above the transition, the uniform susceptibility continues to increase up to fairly high temperatures, with a slope significantly higher than that in the $J_{1a}$-$J_{1b}$-$J_2$ or the $J_1$-$J_2$ model.  (iv) The behavior of the specific heat and the order-parameter at the transition are very close to the corresponding Ising model on the same lattice, once the temperatures are scaled by the peak values. This suggests that the transition is continuous and of second order, belonging to the universality class of the 2D Ising model.

\begin{figure}[t!]
\includegraphics[width=\columnwidth]{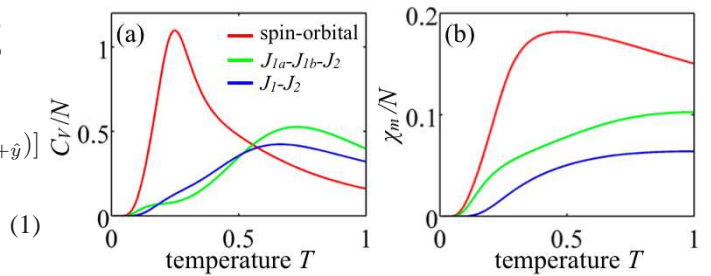}
\caption{
(Color online) Plots for (a) the specific heat $C_V$, and (b) the uniform magnetic susceptibility $\chi_m$ for the three models considered. The temperature $T$ is expressed in terms of $J_1$ (or $J_{1a}$). Compared to the other two spin-only models, there is a sharp peak in $C_V$ and a larger slope in $\chi_m$ in the spin-orbital model.
}\label{fig:CV_UM}
\end{figure}

Fig. \ref{fig:CV_UM} shows the specific heat $C_V$ and uniform magnetic susceptibility $\chi_m$ for the spin-orbital model; for comparison we also plot the same quantities for the spin-$\frac{1}{2}$ $J_{1a}$-$J_{1b}$-$J_2$ (with $J_{1b}=-0.1J_{1a}$, $J_2=0.4J_{1a}$), and $J_1$-$J_2$ (with $J_2=J_1$) models. A main difference in $C_V$ between the spin-orbital and the other two spin-only models is the sharp peak at $T\sim 0.23J_1$, an indication of a phase transition.

For an AF ordered ground state, $\chi_m$ will grow as $T$ increases from $T$=0, and then turn down at some characteristic temperature associated with short-range magnetic order. $\chi_m$ at $T=$0 should have a finite value due to gapless excitations (Goldstone modes) intrinsic to each model in the thermodynamic limit. This is not captured in ED due to finite size effects. Nonetheless, one expects the results to be qualitatively valid near the peak and quantitatively valid above it \cite{marcos}. With our parameters, the energy to flip a spin in the AF ground state is approximately $J_1+J_f+2J_2\approx 2J_1$. Hence the $T=0$ $\chi_m$ should be comparable to that of an isotropic square-lattice Heisenberg model with the same $J_1$ \cite{hamer,notefactor2by3}. One then expects for both the spin-orbital and the $J_{1a}$-$J_{1b}$-$J_{2}$ models an identical susceptibility below $T=0.1 J_1$, with a magnitude of $\sim0.05/J_1$. The sharp difference is the sudden increase in $\chi_m$ between $T=0.1 J_1$ and the phase transition $\sim$ $T=0.2 J_1$.

\begin{figure}[t!]
\includegraphics[width=\columnwidth]{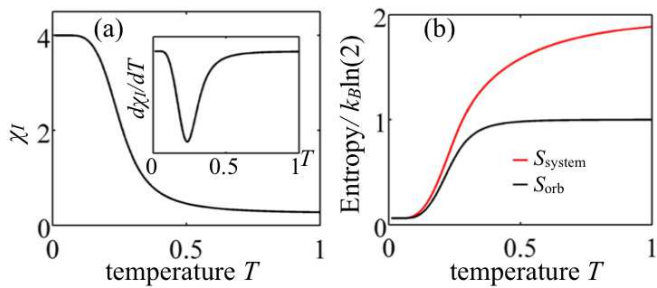}
\caption{
(Color online) (a) The orbital-Ising Susceptibility $\chi_I$ and its derivative with respect to $T$ in the inset. (b) The total system $S_\text{system}$ and orbital $S_\text{orb}$ entropy in the spin-orbital model. In the vicinity of the phase transition $\sim$ $R \ln{2}$ entropy is lost.
}\label{fig:Ising_ent}
\end{figure}

A direct way to locate the orbital ordering transition temperature $T_c$ is through the orbital Ising susceptibility $\chi_I$:
\begin{equation}
\chi_I \equiv \frac{1}{N} \sum_{\alpha} P_\alpha (N_t-N/2)^2.
\end{equation}
The sum on $\alpha$ is over the $2^{16}$ Ising configurations, with $P_\alpha$ the probability of the $\alpha^{th}$ configuration. $N_t\equiv \sum_i n_i$ is the sum of the Ising variables on the lattice tied to the $\alpha^{th}$ Ising configuration. According to the definition, $\chi_I$ is $N/4$ at $T=0$, and monotonically decreases to the configuration averaged value as $T$ increases. The peak in $d\chi_I/dT$ is a measure of $T_c$ which happens at $\sim 0.23J_1$, as indicated by Fig. \ref{fig:Ising_ent}(a).

We can define an orbital entropy $S_\text{orb}\equiv -\frac{1}{N}\sum_\alpha P_\alpha \ln P_\alpha$, which approaches $\ln(2)$ at high temperature. On the other hand, the total system entropy $S_\text{system}$ incorporating both spin and orbital degrees of freedom (obtained by integrating $C_V/T$ with respect to $T$) approaches $\ln(4)$ per site as $T$ increases. Fig. \ref{fig:Ising_ent}(b) indicates that $S_\text{orb}$ is completely exhausted soon after $T_c$, saturating much faster than $S_\text{system}$. We have checked that the behavior of the Ising variables in the spin-orbital model is quantitatively very close to the pure Ising model once the temperatures are scaled according to their corresponding $C_V$ peak values. This suggests that the orbital phase transition is in the 2D Ising universality class where the finite temperature phase transition is continuous and of second order. A more definitive conclusion would require study via other numerical techniques such as quantum Monte Carlo on larger systems. This, however, may face minus sign problems because the spin-orbital model is frustrated.

\begin{figure}[t!]
\includegraphics[width=\columnwidth]{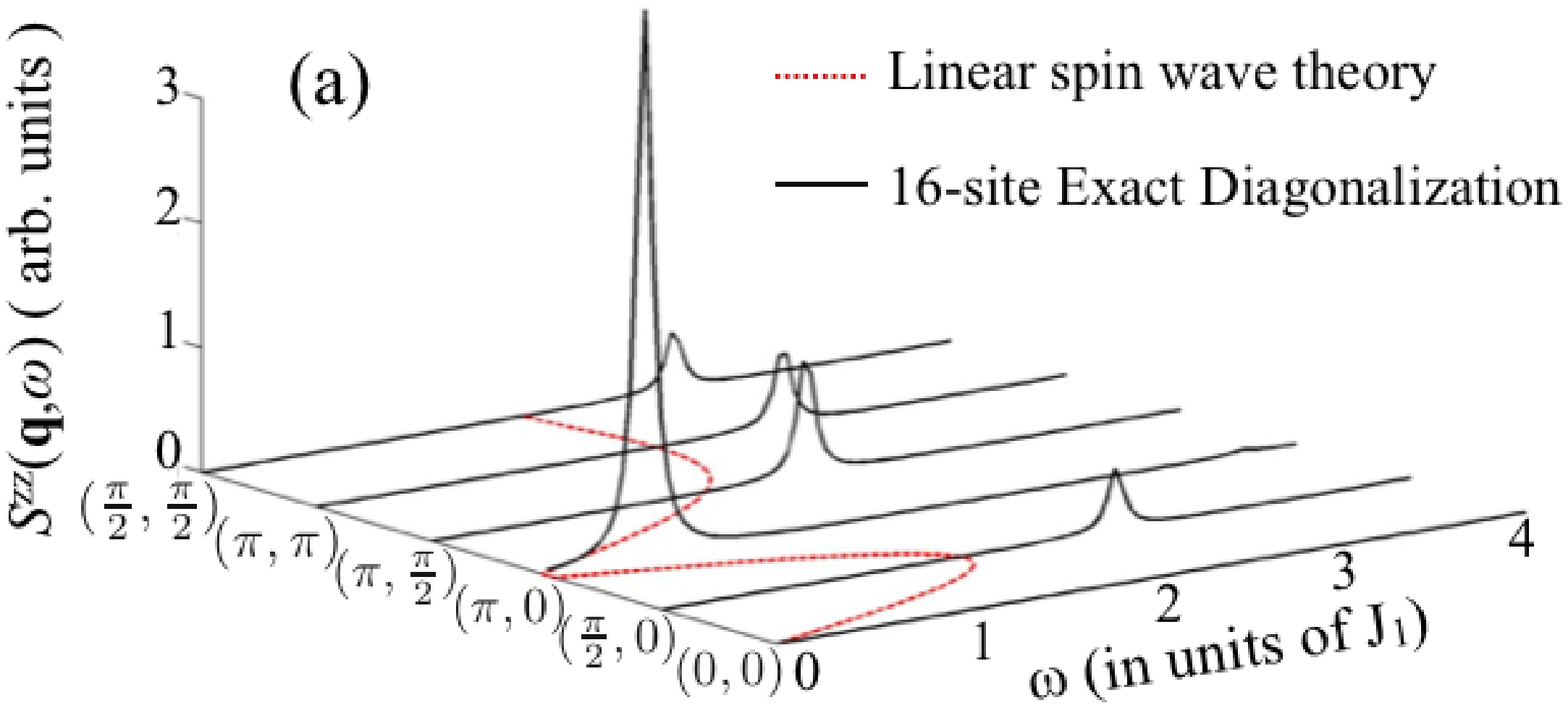}
\includegraphics[width=\columnwidth]{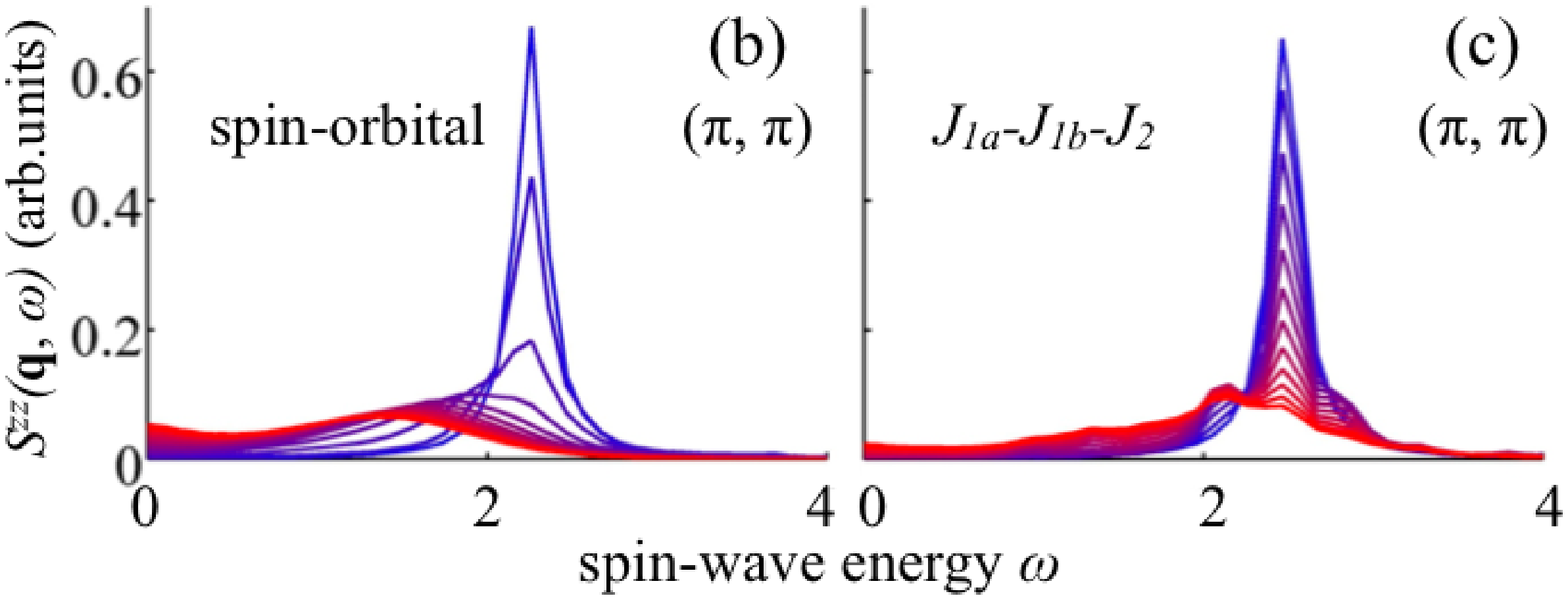}
\includegraphics[width=\columnwidth]{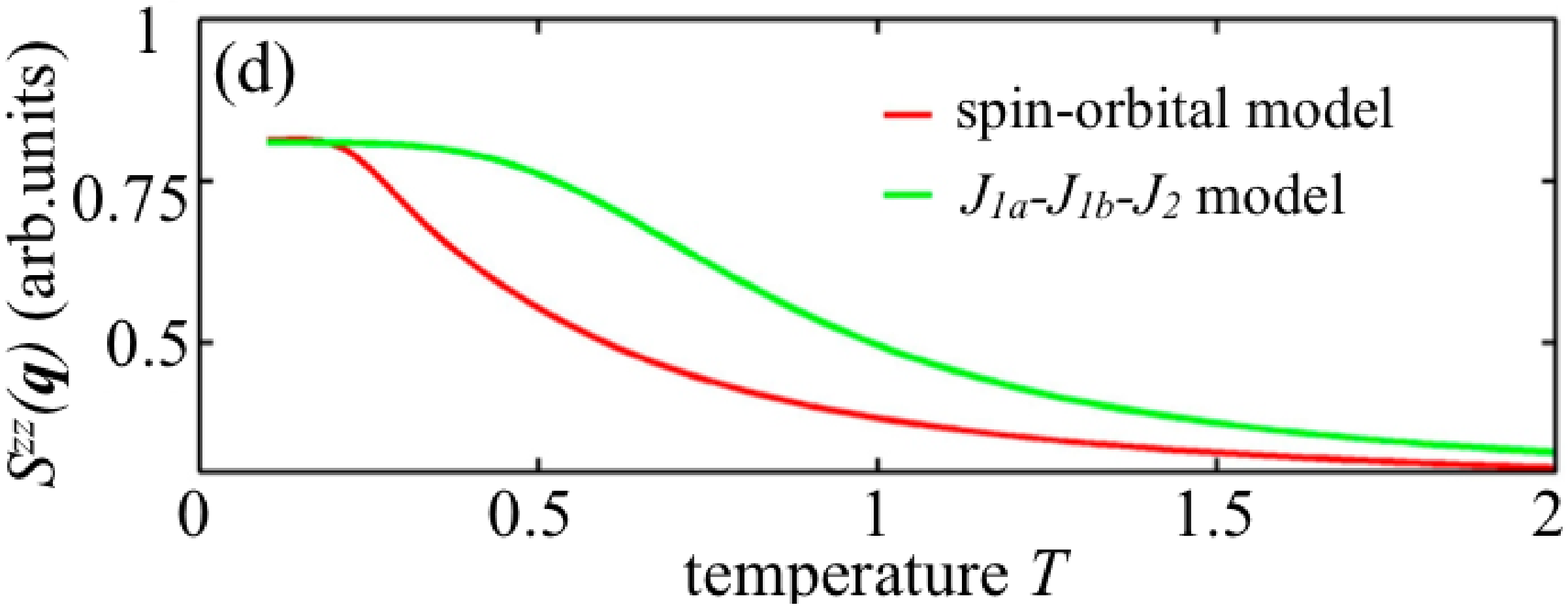}
\caption{
(Color online) (a) $T=0$ $S^{zz}(\mathbf{q},\omega)$ for the spin-orbital/$J_{1a}$-$J_{1b}$-$J_2$ models. (b)-(c) Finite temperature $S^{zz}(\mathbf{q},\omega)$ at $(\pi,\pi)$ obtained from ED for the spin-orbital, and the $J_{1a}$-$J_{1b}$-$J_2$ models, respectively. The temperature goes from $T=0.1J_1$ (the blue curve) to $T=1.0J_1$ (the red curve), with a temperature increment between each curve $\sim0.08J_1$. A highly incoherent spin dynamics is observed in the spin-orbital model. (d) Finite temperature $\omega$-integrated form factor $S^{zz}(\mathbf{q})$ at $(\pi,0)$.
}\label{fig:SQW}
\end{figure}

We next turn our focus to the spin dynamics of the spin-orbital model by studying the dynamic form factor $S^{\alpha\beta}(\mathbf{q},\omega)$, which is the Fourier transform of the spin-spin correlation function $\langle S^\alpha_i(t)\cdot S^\beta_j (t')\rangle$. We calculate $S^{zz}(\mathbf{q},\omega)$ via both ED and linear spin wave theory. Apart from the small energy gap in ED due to finite size effects, the spin wave dispersions obtained from both methods are compatible, see Fig. \ref{fig:SQW} (a). Neutron scattering on Fe-pnictide parent compounds indicates that the magnon energy is a maximum at $(\pi,\pi)$. This behavior, absent in the $J_1$-$J_2$ model, is captured correctly by the $J_{1a}$-$J_{1b}$-$J_2$ model, and hence the $T=0$ spin-orbtial model.

At finite temperatures, orbital fluctuations start to play a role. The spectra in the spin-orbital model broaden much faster then the $J_{1a}$-$J_{1b}$-$J_2$ model and show anomalous shifts to low frequencies. Fig. \ref{fig:SQW} (b) and (c) show the temperature evolution of spin wave at $(\pi,\pi)$ obtained from ED. At temperatures higher than $\sim 0.3J_1$, only incoherent spin waves survive in the spin-orbital model. In contrast, in the $J_{1a}$-$J_{1b}$-$J_2$ model the coherent spin waves persist to a temperature higher than $T\sim 0.6J_1$. This feature can be seen also in the $\omega$-integrated form factors $S^{\alpha\beta} (\mathbf{q}) \equiv \int  d\omega S^{\alpha\beta}(\mathbf{q},\omega)$. For the spin-orbital model, from $T=0$ to $0.4J_1$ the dominant peak at $(\pi,0)$ decreases by 25\% in intensity, while for the $J_{1a}$-$J_{1b}$-$J_2$ model it requires a temperature higher than $T=0.8J_1$ to show a similar reduction, see Fig. \ref{fig:SQW} (d). Thus finite temperature neutron spectra can distinguish these models.

Before we continue to discuss the relevance of this study to the Fe pnictides, a few comments are in order. The issue regarding the correlation strength in the Fe-pnictide materials is controversial and currently under debate. Recently, x-ray absorption data on several Fe-pnictide compounds revealed that the on-site Coulomb repulsion was smaller than the bandwidth, but also found a substantial Hund's coupling $J_H=0.8$ eV between the Fe 3$d$ orbitals \cite{seminal-work}. Moreover, there is no particular energy scale above which damped spin waves are found \cite{Dai}; this absence of a Stoner decay strongly favors a picture based on localized moments. 

In many regards, the pnictides are schizophrenic, having aspects such as metallicity and strong covalency where correlations play a minor role \cite{seminal-work, correlations}, and anti-ferromagnetism and local properties which derive directly from the strength of the Hund's coupling. Therefore a model based on local moments which takes aim at the magnetic properties of the pnictides and other transition metal oxide is completely in line with the findings in Ref. \cite{seminal-work}, and more recently with observations from optical conductivity measurements \cite{mazin_OC}. Our model focuses on a subset of localized orbitals in connection with magnetism. but neglects the fact that the 5 Fe $d$ orbitals in Fe pnictides are not strongly crystal field split.

Certain details of the model can be modified easily without changing the essential features. For example, there can be a direct coupling between the Ising variables reflecting lattice effects and quadrupolar couplings. In addition, the local environment of As positions could modify local field screening and exchanges. These changes will alter the orbital gap and transition temperature, but not the overall picture. We will make our comparisons only in semi-quantitative terms.

With this in mind, the spin-orbital model captures many features of the uniform susceptibility in the pnictides \cite{DH_sus}. For example, in BaFe$_2$As$_2$ the susceptibility in emu/mole is $0.6 \times 10^{-3}$ at $T=0$; it sharply increases near $T=150 K$ to about $0.9\times 10^{-3}$ and then continues to increase linearly to about $1.5\times 10^{-3}$ at $T=600 K$. If part of the susceptibility is a weakly temperature dependent Pauli term, this implies an increase by a factor of about 3 between $T=0$ and $T=600 K$. Our finite-size calculations can be converted \cite{coldea} to emu/mole by multiplying the susceptibility by a factor $(8*g^2*0.0938)/J_1$, where $J_1$ is in Kelvin and a factor of 2 comes from the 2 Fe atoms per mole of the material. With $J_1\approx 1000 K$, this gives a susceptibility in cgs units of order $10^{-3}$ at $500 K$, which comes down by about a factor of $3-4$ by $T=0$ including a sharp drop below the transition.

In the 1111 pnictide family, two phase transitions at nearby temperatures have been reported, a structural transition at higher $T$ and a magnetic transition at lower $T$. In contrast, only one simultaneous structural and magnetic transition is found in the 122 family. This is explained naturally in terms of 3D couplings. Orbital ordering driven by magnetism requires the prior development of short-range spin order. A 3D system with strong inter-planar coupling would therefore lead to simultaneous spin and orbital order. In contrast, for a weakly inter-planar coupled 2D system orbitals order when short-range spin order develops, but spins only order on the scale of inter-planar couplings. This therefore leads to two separate transitions. These observations are indeed consistent with the two different pnictide families. This aspect also has been suggested for $J_1$-$J_2$ Heisenberg models \cite{strong}. However, one important difference is that in the spin-orbital model $\sim$ $R \ln{2}$ entropy is lost in the vicinity of the transition; it is likely significantly smaller in the $J_1$-$J_2$ models. It is noted that in the Fe$_{1+y}$Se$_x$Te$_{1-x}$ systems a comparable amount of entropy change is found near the AF transition \cite{entropy-change}.

In summary, we have studied a model system that captures the physics of coupled spin and orbital degrees of freedom. Such a system apparently undergoes a continuous, second-order phase transition to an orbitally ordered state at a temperature set by short-range magnetic order. The onset of orbital excitations and fluctuations cause a highly incoherent spin dynamics, leading to a sharp increase in uniform magnetic susceptibility. The susceptibility continues to increase up to fairly high temperature above the phase transition, with a large slope comparable to those observed in the pnictides. Our calculations of dynamic structure factors at finite temperatures serve as clear predictions of the spin-orbital model, that can be tested by further experiments. In addition to the pnictides, the model should be generally applicable to other systems with orbital degeneracy, with the strengths/signs of the exchange constants dependent on the microscopic details.

\begin{acknowledgments}
We acknowledge useful discussion with J.-H. Chu, J. Maciejko, S. Johnston, A. P. Sorini, H. Yao, E. Berg, W. A. Harrison, W. Pickett, S. Savrasov and T. Yildirim. This work was supported by the Office of Science of the U.S. Department of Energy (DOE) under Contract No. DE-AC02-76SF00515. This research used resources of the National Energy Research Scientific Computing Center, supported by DOE under Contract No. DE-AC02-05CH11231.
\end{acknowledgments}


\begin{thebibliography}{arprev}

\bibitem{Kugel}
K. I. Kugel and D. I. Khomskii, Sov. Phys. Usp. \textbf{136}, 621 (1982).

\bibitem{Tokura}
Y. Tokura and N. Nagaosa, Science \textbf{288}, 462 (2000).

\bibitem{pnictide_family}
Y. Kamihara \emph{et al.}, J. Am. Chem. Soc. \textbf{130}, 3296 (2008);
M. Rotter \emph{et al.}, Phys. Rev. B \textbf{78}, 020503 (R) (2008);
X.C.Wang \emph{et al.}, Solid. State. Comm. \textbf{148}, 538 (2008);
F.-C. Hsu \emph{et al.}, Proc. Nat. Acad. Sci.(USA) \textbf{105}, 14262 (2008). 

\bibitem{Dai}
Clarina de la Cruz \emph{et al.}, Nature \textbf{453}, 899 (2008).

\bibitem{weak}
V. Cvetkovic and Z. Tesanovic, Europhys. Lett. 85, 37002 (2009);
A. V. Chubukov \emph{et al}., Phys. Rev. B \textbf{78}, 134512 (2008);
S. Raghu \emph{et al}., Phys. Rev. B \textbf{77}, 220503 (R) (2008);
Y. Ran \emph{et al}., Phys. Rev. B \textbf{79}, 014505 (2009);
K. Seo \emph{et al}., Phys. Rev. B \textbf{79}, 235207 (2009).

\bibitem{strong} 
K. Haule \emph{et al}., Phys. Rev. Lett. \textbf{100}, 226402 (2008);
Q. Si and E. Abrahams, Phys. Rev. Lett. \textbf{101}, 076401 (2008);
C. Fang \emph{et al}., Phys. Rev. B \textbf{77}, 224509 (2008);
C. Xu \emph{et al}., PRB \textbf{78}, 020501 (2008);
G. Baskaran, J. Phys. Soc. Jpn. \textbf{77}, 113713 (2008);
G. S. Uhrig \emph{et al}., Phys. Rev. B \textbf{79}, 092416 (2009);
E. Berg \emph{et al.}, arXiv:0905.1096.

\bibitem{J1_J2}
P. Chandra \emph{et al.}, Phys. Rev. Lett. \textbf{64} 88 (1990).

\bibitem{savrasov}
Z. P. Yin \emph{et al}., Phys. Rev. Lett. \textbf{101}, 047001 (2008);
M. J. Han \emph{et al}., Phys. Rev. Lett. \textbf{102}, 107003 (2009);
T. Yildirim, Phys. Rev. Lett. \textbf{101}, 057010 (2008).

\bibitem{Neutron_SW}
Jun Zhao \emph{et al}., 	arXiv:0903.2686.

\bibitem{Jeroen_OO}
Frank Kr\"uger \emph{et al.}, Phys. Rev. B \textbf{79}, 054504 (2009).

\bibitem{Rajiv_OO}
Rajiv R. P. Singh, arXiv:0903.4408.

\bibitem{edge-sharing}
Y. Mizuno \emph{et al}., Phys. Rev. B \textbf{57}, 5326 (1998);
F. Vernay \emph{et al}., Phys. Rev. B \textbf{77}, 104519 (2008).

\bibitem{Double_Exchange}
Clarence Zener, Phys. Rev. \textbf{82}, 403 (1951); P. W. Anderson and H. Hasegawa, Phys. Rev. \textbf{100}, 675 (1955).

\bibitem{other_OO}
T. Yildirim, arXiv: 0902.3462;
Weicheng Lv \emph{et al}.,  arXiv: 0905.1704;
Ari M. Turner \emph{et al.}, arXiv: 0905.3782.

\bibitem{savrasov-privatecomm}
S. Savrasov, private communication.

\bibitem{WeiKu_OO}
Chi-Cheng Lee \emph{et al.}, arXiv:0905.2957.

\bibitem{Dagotto}
E. Dagotto, Rev. Mod. Phys. \textbf{66}, 763 (1994).

\bibitem{marcos}
M. Rigol \emph{et al}., Phys. Rev. E \textbf{75}, 061118 (2007).

\bibitem{hamer}
Z. Weihong \emph{et al.}, Phys. Rev. B \textbf{43}, 8321 (1991).

\bibitem{notefactor2by3}
The $T\to 0$ limit of the susceptibility ie related to $T=0$ transverse susceptibility by a factor of 2/3.

\bibitem{seminal-work}
W. L. Yang \emph{et al.}, Phys. Rev. B \textbf{80}, 014508 (2009);
V. I. Anisimov \emph{et al}., Physica C \textbf{469}, 442 (2009).

\bibitem{correlations}
S. L. Skornyakov \emph{et al.}, arXiv:0906.3218;
M. Aichhorn \emph{et al.}, arXiv:0906.3735.

\bibitem{mazin_OC}
S. J. Moon \emph{et al.}, arXiv:0909.3352.

\bibitem{DH_sus}
G. Wu \emph{et al}, J. Phys. Condens. Matter \textbf{20}, 422201 (2008);
X. F. Want, New J. Phys. 11 (2009) 045003;
G. M. Zhang \emph{et al}., Euro. Phys. Lett. \textbf{86}, 37006 (2009).

\bibitem{coldea}
W. Zheng \emph{et al}., Phys. Rev. B \textbf{71}, 134422 (2005).

\bibitem{entropy-change}
S. Li \emph{et al}., Phys. Rev. B \textbf{79}, 054503 (2009).

\end{thebibliography}
\end{document}